%% file: paper.tex
\documentclass{WileyMSP-template}

\usepackage[T1]{fontenc}             
\usepackage[utf8]{inputenc}          
\usepackage[english]{babel}          

\usepackage{amsmath}                 
\usepackage{amssymb}                 
\usepackage{mathrsfs}                
\usepackage{mathtools}               
\usepackage{dsfont}                  

\usepackage{microtype}

\usepackage{physics}

\usepackage{tikz}
\usepackage{standalone}
\usepackage{contour}
\usetikzlibrary{3d}
\usetikzlibrary{backgrounds}
\usetikzlibrary{decorations.pathreplacing}

\usepackage[unicode]{hyperref}  
\usepackage[noabbrev,capitalise]{cleveref} 

\graphicspath{{figures/}}
\begin{document}

\pagestyle{fancy}

\justifying
\setlength{\parindent}{0pt}

\input{commands}

\title{Spin-polarization and resonant states in electronic conduction through a correlated magnetic layer}

\maketitle

\author{Andreas Weh*}
\author{Wilhelm H. Appelt}
\author{Andreas \"Ostlin}
\author{Liviu Chioncel}
\author{Ulrich Eckern}

\begin{affiliations}
A.\ Weh, W.\ H.\ Appelt, U.\ Eckern\\
Institute of Physics, University of Augsburg, 86135 Augsburg, Germany\\
Email Address: andreas.weh@physik.uni-augsburg.de

A.\ \"Ostlin, L.\ Chioncel\\
Center for Electronic Correlations and Magnetism
\& Augsburg Center for Innovative Technologies,
Institute of Physics, University of Augsburg, 86135 Augsburg, Germany

\end{affiliations}


\begin{abstract}
The transmission through a magnetic layer of correlated electrons sandwiched between non-interacting normal-metal leads is studied within model calculations.
We consider the linear regime in the framework of the Meir-Wingreen formalism, according to which the transmission can be interpreted as the overlap of the spectral function of the surface layer of the leads with that of the central region.
By analyzing these spectral functions, we show that a change of the coupling parameter between the leads and the central region significantly and non-trivially affects the conductance.
The role of band structure effects for the transmission is clarified.
For a strong coupling between the leads and the central layer, high-intensity localized states are formed outside the overlapping bands, while for weaker coupling this high-intensity spectral weight is formed within the leads' continuum band around the Fermi energy.
A local Coulomb interaction in the central region modifies the high-intensity states, and hence the transmission.
For the present setup, the major effect of the local interaction consists in shifts of the band structure, since any sharp features are weakened due to the macroscopic extension of the configuration in the directions perpendicular to the transport direction.
\end{abstract}

\setlength{\parindent}{14pt}

\section{Introduction}\label{sec:intro}
The transport properties of inhomogeneous electronic systems, including charge, spin and heat flow, have been intensely studied for several years~\cite{heath2009,wolf2001,dubi2011}, in particular, in view of the enormous application potential of such devices.
As a specific example, we wish to mention experimental investigations of the tunneling magnetoresistance effect in Fe/MgO/Fe magnetic tunnel junctions, which were created by depositing MgO epitaxially between Fe electrodes~\cite{mo.ki.95}.
The quantum transport properties of such structures can be described employing
model Hamiltonians~\cite{ma.um.01}, or density functional theory (DFT)~\cite{bu.zh.01}.

By both these approaches the spin dependence of tunneling can be explained by the nature of the Bloch states in the bulk ferromagnetic leads that couple with evanescent states of the complex band structure of the insulator~\cite{ma.zh.99,ma.pa.00}.
Depending on the strength of this coupling, the majority/minority conductance is expected to be dominant, thus determining the positive/negative spin polarization of the conductance.
The interface bonding between leads and the insulating region is considered to be responsible for the formation of interface states~\cite{ts.pe.97}, which significantly affect the conduction spin polarization;
their existence was experimentally confirmed~\cite{ti.fa.04} through spin-dependent tunneling measurements.  
Interfaces between ferromagnetic and non-magnetic metals have also been of recent interest since they are important for current-perpendicular-to-plane giant magnetoresistance (CPP-GMR) experiments.
 Initially the magnetic and the non-magnetic spacer layers were mainly composed of transition metals
 like Co/Ag, as in the pioneering work of Ref.~\cite{pr.le.91}; however,
 recently more complex Heusler alloys~\cite{ba.lu.12,ro.pr.18} have also been investigated due to their high spin polarization.

In the theory of the CPP-GMR~\cite{va.fe.93}, spin-asymmetric scattering plays an important role:
the larger the asymmetry, the larger the possible spin polarization of the current.
In order to achieve a high spin asymmetry at the interface, it was proposed \cite{ni.ko.09} that materials should be employed where the electronic band structure of the majority spin of the ferromagnetic layer matches as closely as possible the band structure of the non-magnetic layer. 
At the same time, the matching between band structures should be poor for the minority spin channel. 
Current theories of spin-dependent tunneling hardly emphasize electronic correlations, neither in the leads nor in the scattering region. In this context,
we have studied within the framework of DFT the ballistic conduction through transition metal heterostructures~\cite{ch.mo.15}, and more recently Heusler based systems such as Au/NiMnSb/Au~\cite{mo.ap.17}. For these system,
we modeled electronic correlations for the Mn atoms via the Hubbard model, which was solved within dynamical mean-field theory (DMFT)~\cite{me.vo.89,ge.ko.96,ko.vo.04}.
We obtained a significant reduction of the spin polarization in the density of states which is not apparent in the spin polarization of the conduction electron transmission, and
concluded that the interface states hybridized with the many-body induced states are localized.
In the present work, we revisit our recent studies using a simplified single-band model Hamiltonian on the one hand, but a more advanced many-body solver to treat electronic correlations on the other hand.  
Within this model, we discuss the nature of the electronic states in the interacting region
which determine the transmission.
We show that a modification of the coupling between leads and the central region, as well as electronic correlations, may strongly affect the spectral functions and hence the conductance.
 Our focus is on a ferromagnetic metal as the central layer, coupled to non-interacting normal-metal leads.
We show, in particular, that electronic interactions may enhance the spin polarization of the spectral function in the interacting region, as well as of the transmission through the entire system. 
The paper is organized as follows. In the next section, \Cref{sec:model}, we present the tight-binding model to be employed in this work. \Cref{sec:results} is devoted to our results,
and \Cref{sec:conclusion} to the summary and conclusions.

\section{Generic transport model}\label{sec:model}
The electronic transport through a device can be conveniently addressed by applying scattering theory, which was pioneered by Landauer~\cite{land.57b,land.88} and Büttiker~\cite{butt.86,butt.88}, and worked out in detail by Meir and Wingreen~\cite{me.wi.92}.
In this approach (see, e.g., Ref.~\cite{ra.sm.86}), one typically considers a mesoscopic system (say, a molecule or a quantum dot) which is coupled via ideal leads which act as charge reservoirs which are so large that they can be described by equilibrium distributions. 
Hence the corresponding left (\(L\)) and right (\(R\)) leads are characterized by the equilibrium Fermi distribution functions \(f_{L/R}(\omega)\).
For the derivation, one assumes that in the infinite past 
the various subsystems (microscopic contact, or central region; and the leads) are separated and in equilibrium, albeit at their respective chemical potentials and temperatures.
The couplings are then turned on adiabatically, assuming time-reversal invariance~\cite{ca.co.no.71,ca.co.le.71}.
In the following, we apply the Meir-Wingreen approach~\cite{me.wi.92} to our heterostructure setup, in which electronic correlations are considered in the scattering region only, i.e., in the central layer. Note that this layer, and all others, are macroscopic in the directions perpendicular to the transport direction.

Recently we have described~\cite{ch.mo.15,mo.ap.17} how to take into account local interactions when computing the transmission of correlated heterostructures.
In comparison to the previous DFT+DMFT approach, in the present tight-binding Hamiltonian the computation of the transmission becomes even more transparent.
In particular, using the Meir-Wingreen formalism, we can replace the scattering region Green's function directly by its interacting counterpart, in which electronic correlations are taken into account by the local self-energy \(\Sigma(\omega)\).
This latter quantity is computed using the recently developed fork tensor-product states (FTPS) solver~\cite{ba.zi.17};
we used this method recently in order to describe the spectral properties of heterostructures containing half-metals~\cite{we.ot.20}.

As is well known, the Meir-Wingreen approach~\cite{me.wi.92} can be considerably simplified for non-interacting systems, or when the coupling matrices are proportional to each other. We emphasize that the latter is fulfilled for our setup, as long as the central region consists of just one interacting layer. On the other hand, for two or more interacting layers, the lesser Green's function will be needed.

\begin{figure}[htbp]
  \centering
  \input{tikz/structure.tikz.tex}
  \caption{%
    Schematic graph of the model setup. The layers are labeled with $l$, such that $l<0$ corresponds to the left, and $l>0$ to the right lead. The
    central layer is $l=0$.
  }\label{fig:model_structure}
\end{figure}
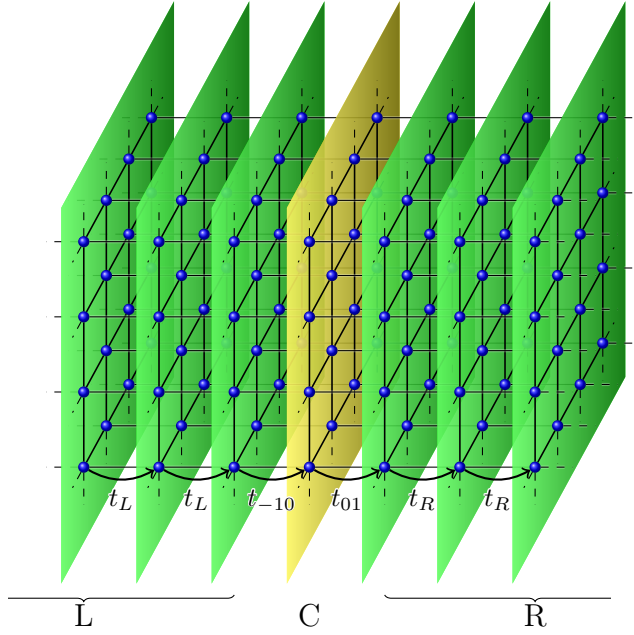

\Cref{fig:model_structure} shows the geometry of the system: non-interacting leads, left ($L$) and right ($R$), separated by the central region ($C$).
Both leads consist of a semi-infinite stack of square-lattice planes.
The hopping amplitude between the layers in the left (right) lead is \(t_{L}\) (\(t_{R}\)), and
the on-site energy is \(\epsilon_{L \sigma}\) (\(\epsilon_{R \sigma}\));
the electron dispersion within the layers is of Bloch type because of two-dimensional translation invariance.
The transverse (i.e., parallel to the layers) wave vector, \(\kp = (k_{x}, k_{y})\), is a good quantum number, and the in-plane dispersion of the electrons is denoted as \(\epskp\). 
The complete Hamiltonian thus can be written as
\(\hat{H} = \hat{H}_{L} + \hat{H}_{C} + \hat{H}_{R}\), where $\hat{H}_{C}$ contains the coupling between leads and central region.
To be definite, the Hamiltonians for the semi-infinite left and right leads read:
\begin{align}
  \hat{H}_{L}
  &= \sum_{l< 0 \kp \sigma} (\epsilon_{L \sigma} + \epskp) \hat{n}_{l\kp \sigma}
  - \sum_{l<0 \kp \sigma} (t_{L} \ocd_{l-1 \kp \sigma} \oc_{l\kp \sigma} + t_{L}^{*} \ocd_{l \kp \sigma} \oc_{l-1 \kp \sigma}),
  \\
  \hat{H}_{R}
  &= \sum_{l> 0 \kp \sigma} (\epsilon_{R \sigma} + \epskp) \hat{n}_{l\kp \sigma}
  - \sum_{l>0 \kp \sigma} (t_{R} \ocd_{l \kp \sigma} \oc_{l+1\kp \sigma} + t_{R}^{*} \ocd_{l+1 \kp \sigma} \oc_{l \kp \sigma}),
\end{align}
where \(\sigma = \pm \frac{1}{2}\) labels the spin.
In the following, we consider the leads
to be identical, but keep the notation ``$L$'' which henceforth refers to ``lead''.
For the central region, just one layer, we have:
\begin{equation}\label{eq:H_C}
  \hat{H}_{C}
  = -\sum_{\kp \sigma} (t_{-10} \ocd_{-1\kp \sigma}\oc_{0\kp \sigma} + \mathrm{h.c.})
  + \sum_{\kp \sigma} (\epsilon_{0 \sigma} + \epskp) \hat{n}_{0\kp \sigma} + U_{0} \sum_{i} \hat{n}_{0i \downarrow} \hat{n}_{0i \uparrow}
  - \sum_{\kp \sigma} (t_{01} \ocd_{0\kp \sigma}\oc_{1\kp \sigma} + \mathrm{h.c.})
\end{equation}
The amplitudes describing the hybridization between the central region and the leads, \(t_{-10}\) and \(t_{01}\), are chosen to be real and positive, and equal to each other.
The hopping amplitudes in the transport direction are also assumed to be real, \(t_{L} = t_{L}^{*}\), without loss of generality. In addition,
\(\epsilon_{0 \sigma} = \epsilon_{0} + \sigma h_{0}\) is the on-site energy, $h_0$ the magnetic splitting,
and \(U_{0}\) the on-site Hubbard interaction in the central layer.
Since the central region is just a \emph{single} layer,
the effect of the leads is characterized by the \emph{scalar} level-width function
\begin{equation}
  \Gamma^{L}_{\sigma}(\omega, \epskp)
  = -2 \abs{t_{01}}^{2} \Im \gsurf_{L}(\omega^{+} - \epsilon_{L \sigma} - \epskp),
\end{equation}
where \(\omega^{+} = \omega + i0^{+}\), with the leads' surface Green's function
\begin{equation}\label{eq:surf_gf}
  \gsurf_{L}(z)
  = \frac{z}{2\abs{t_{L}}^{2}} \left(1 - \sqrt{1 - {\left(\frac{2\abs{t_{L}}}{z}\right)}^{2}}\right).
\end{equation}

Applying the Meir-Wingreen formalism~\cite{me.wi.92} to the present, highly symmetric heterostructure, we obtain a particularly simple expression for the charge current perpendicular to the layers: 
\begin{equation}
  J
  = - \frac{e}{h} N_{\parallel} \sum_{\sigma}\int \mathrm{d} \omega \left[f_{L}(\omega) - f_{R}(\omega)\right]
  \int \mathrm{d} \epsp \rhop(\epsp) \Gamma^{L}_{\sigma}(\omega, \epsp)\Im G_{00 \sigma}(\omega^{+}, \epsp) ,
\end{equation}
where \(G_{00 \sigma}\) is the Green's function of the central region, \(G_{00 \sigma} (z, \epskp) = \gfop{\oc_{0\kp \sigma}}{\ocd_{0\kp \sigma}}(z)\),
and \(N_{\parallel}\) the number of sites within a layer perpendicular to the transport direction.
The layer density of states is given by
\begin{equation}
  \rhop(\epsp)
  = \frac{1}{N_{\parallel}}\sum_{\kp} \delta(\epsp - \epskp)
  = \frac{2}{\pi^2 D} K(1 - {\epsp^{2}}/{D^{2}}),
\end{equation}
where the last equality is valid for $\abs{\epsp} \le D$.
For the square lattice~\cite{econ.06}, it can be written in terms of the complete elliptic integral of the first kind:
\begin{equation}
  K(m) = \int_{0}^{\pi /2} \mathrm{d}t {[1-m\sin^{2}(t)]}^{-1/2}.
\end{equation}
From the above quantities, the normalized transmission for the spin channel $\sigma$ can be computed as follows:
\begin{equation}\label{eq:T_omega}
\begin{aligned}
  T_{\sigma}(\omega)
  &= -\int \mathrm{d} \epsp \rhop(\epsp) \Gamma^{L}_{\sigma}(\omega, \epsp) \Im G_{00 \sigma}(\omega^{+}, \epsp)
  \\
  &= 2 \abs{t_{01}}^{2} \int \mathrm{d} \epsp \rhop(\epsp)
  \Im \gsurf_{L}(\omega^{+} - \epsp - \epsilon_{L \sigma})
  \Im G_{00 \sigma}(\omega^{+}, \epsp).
\end{aligned}
\end{equation}
Within dynamical mean-field theory (DMFT), the local self-energy of the central region, \(\Sigma_{\sigma}(z)\), is included in the central region Green's function:
\begin{equation}\label{eq:Gf_center}
  G_{00 \sigma}(z, \epsp)
  = \frac{1}
  {z - \epsp - \epsilon_{0 \sigma} - \Sigma_{\sigma}(z) - 2\abs{t_{01}}^{2} \gsurf_{L}(z - \epsilon_{L \sigma} - \epsp)}.
\end{equation}
In the next section, we discuss the behavior of the spin dependency of the spectral functions, and contrast it with the spin-dependent transmissions, when varying the hopping to/from the central region, as well as the strength of the local interaction on the central layer.

\section{Results}\label{sec:results}
We consider the setup as shown in \cref{fig:model_structure}.
As discussed above, our model consists of non-magnetic (non-spin-polarized) metallic leads in contact with a single layer of a ferromagnetic metal.
The Hamiltonian describing the leads (at half-filling) is specified by the on-site energies, \(\epsilon_{L \sigma} = 0\), and the electrons' hopping matrix elements \(t_{L}\) in the direction of transport. The latter are fixed at \(t_{L}=0.25 D\), where \(D\) denotes the parallel half-bandwidth. In addition, the square-lattice parallel hopping matrix elements, \(t_{\parallel}\), are assumed to have the same value.
From now on, \(D\) (\(=4t_{\parallel}\)) will be our energy unit, i.e., formally \(D=1\), and \(t_{L}=t_{\parallel}=0.25\).
According to \cref{eq:T_omega},
the transmission is determined by the product of the surface spectral functions of the uncoupled leads and that of the central region.
The surface Green's function is given above, \cref{eq:surf_gf}.
The corresponding spectral function
\begin{equation}
  A_{L}^{s}(\omega - \epsp)
  = -\frac{1}{\pi}\Im\gsurf(\omega^{+} - \epsp)
\end{equation}
has a semi-circular shape and vanishes at the band edges, \(\pm 2\abs{t_{L}} = \pm 0.5\).
The lead spectral function \(A_{L}^{s}\) does not depend on the parameters of the central layer,
hence our focus is on the spectral function of the central region.
First, we present our results for the non-interacting case, i.e., we discuss how the
spectral function of the central region and the transmission depend
on the parameters on-site energy (\(\epsilon_{0}\)) and coupling (\(t_{01}\)), cf.~\cref{fig:m-ferro-m_aweps_tcXX,fig:trans_U0_resonance,fig:m-ferro-m_U0}.
In the second part of this section, we discuss the modifications induced by a local interaction (\(U_0\)) within the central layer:
We vary \(t_{01}\) for fixed \(U_0\), see \cref{fig:aweps_tcXX_U2}, as well as \(U_0\) for fixed \(t_{01}\), see \cref{fig:pol_Aw_UX}.

\subsection{Non-interacting central layer} 
\begin{figure*}[htbp]
  \centering
  \includegraphics{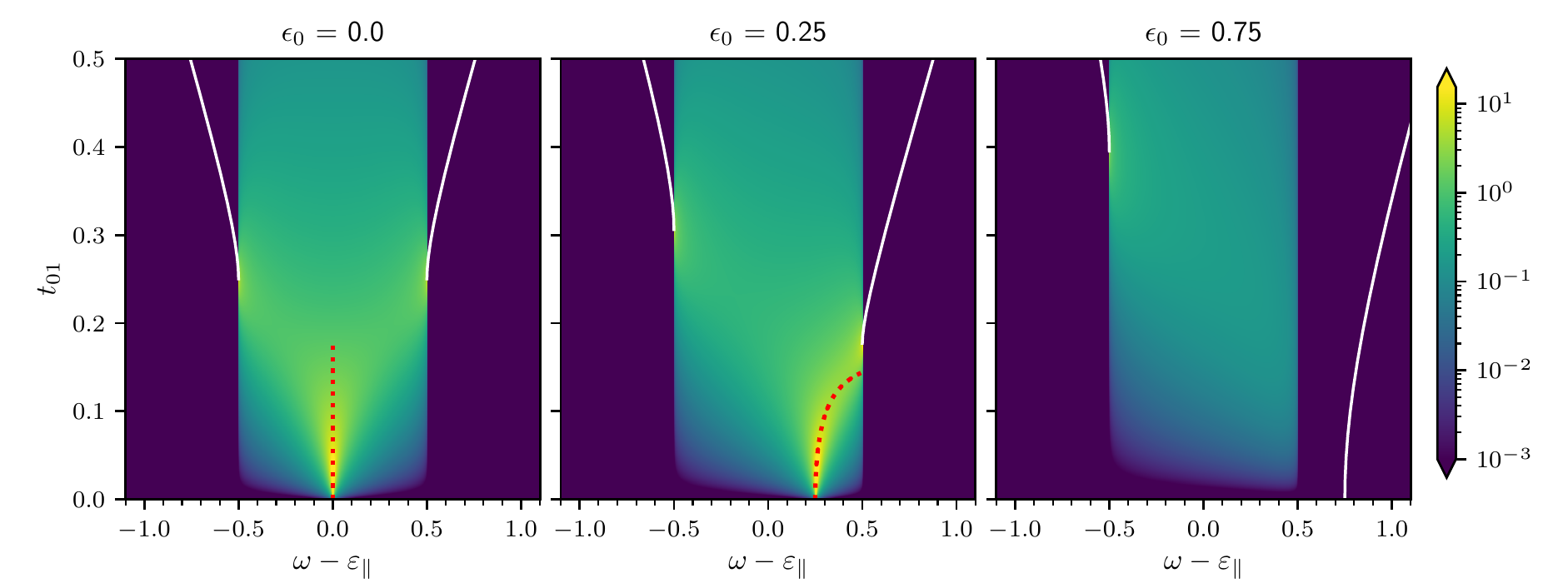}
  \caption{
    Dependence of the spectral function of the central region, \(A_{00\sigma}(\omega, \epsp)\), on \(t_{01}\) for a selection of \(\epsilon_0\).
    The parameters are \(\epsilon_{L \sigma}=0\), \(t_{L} = 0.25\), and \(U_{0} = 0\).
    For a non-interacting system, the spectral function is a function of \(\omega - \epsp\) only: \(A_{00 \sigma}(\omega, \epsp) = A_{00 \sigma}(\omega - \epsp)\).
    The white line indicates the position of the delta peak outside the band, 
    \(\abs{\omega - \epsp} > 2t_L\), while
    the red-dotted line is determined by the real-part of \(\xi_\sigma\) within the band \(\abs{\omega-\epsp} < 2t_L\), \(\omega - \epsp = \Re \xi_\sigma\), 
    resulting in an enhanced spectral intensity; cf.~\Cref{eq:delta-peak-solution}.
  }\label{fig:m-ferro-m_aweps_tcXX}
\end{figure*}

In the non-interacting case, the Green's function depends only on the difference between frequency and dispersion, \(G_{00 \sigma}(\omega, \epskp) = G_{00 \sigma} (\omega - \epskp)\).
The corresponding spectral function is
\begin{equation}\label{eq:A_00_leds}
  A_{00 \sigma}(\omega - \epsp)
  = -\frac{1}{\pi} \Im \frac{1}{\omega^{+} - \epsp - \epsilon_{0 \sigma} - 2 \abs{t_{01}}^{2} g^{s}_{L}(\omega^{+} - \epsp)}.
\end{equation}
In addition, the spin directions are decoupled.
We define \(\xi_\sigma\) as the (generally complex) root of the denominator of the r.h.s.\ of \Cref{eq:A_00_leds}, i.e.,
\begin{equation}
  \xi_\sigma - \epsilon_{0 \sigma} - 2 \abs{t_{01}}^{2} g^{s}_{L}(\xi_\sigma)
  =0,
\end{equation}
which leads to
\begin{equation}\label{eq:delta-peak-equation}
  (1-2r) \xi^{2}_\sigma - 2(1-r) \epsilon_{0 \sigma} \xi_{\sigma} + \epsilon^{2}_{0 \sigma} + 4 r^{2} \abs{t_{L}}^{2}
  = 0,
\end{equation}
where \(r = \abs{t_{01}/t_L}^2\) characterizes the 
hopping to the central region relative to the hopping in the leads.
First we consider energies outside of the lead band, \(\abs{\omega - \epsp} > 2\abs{t_{L}} = 0.5\). Then
the lead Green's function \(g^{s}_{L}\) is real, implying that the spectral function is a sum over delta functions:
\begin{equation}
  A_{00 \sigma}(\omega - \epsp)
  = \sum_{\xi_\sigma} \delta(\omega - \epsp - \xi_{\sigma}).
\end{equation} 
The corresponding \(\xi_\sigma\)-solutions are
\begin{equation}\label{eq:delta-peak-solution}
  \xi^{\pm}_{\sigma} 
  = \frac{(1-r) \epsilon_{0\sigma} \pm r \sqrt{\epsilon_{0 \sigma}^{2} - 4(1-2r)\abs{t_{L}}^{2}}}{1-2r}.
\end{equation}
Note that this expression also contains spurious solutions belonging to the unphysical branch of the square root in \(g^{s}_{L}(z)\).
On the other hand, inside the lead band, \(\abs{\omega - \epsp} < 2 \abs{t_{L}} = 0.5\),
the denominator has an imaginary part, yielding the spectral function
\begin{equation}
  A_{00 \sigma}
  = \frac{1}{\pi} \frac{r (\omega - \epsp) \sqrt{{\left(\frac{2 \abs{t_{L}}}{\omega - \epsp}\right)}^{2} -1}}%
  {(1-2r) {(\omega - \epsp)}^{2} - 2(1-r) (\omega - \epsp) \epsilon_{0 \sigma} + \epsilon_{0 \sigma}^{2} + 4r^{2} \abs{t_{L}}^{2}}.
\end{equation}
Hence within the band we do not get any divergences.
The real part of the roots, \cref{eq:delta-peak-solution}, indicates resonances of increased amplitude in the spectrum.
The results are summarized in \Cref{fig:m-ferro-m_aweps_tcXX} which shows the spectral function, \cref{eq:A_00_leds},
as a function of the energy and the coupling, \(t_{01}\). 
Note that in this representation the spectral function does not depend on the lattice structure of the layers, 
as the horizontal axis refers to \(\omega - \epsp\).

The representative feature of the spectral function is a continuum band in the energy range of \([-2t_{L}, +2t_{L}] = [-0.5, +0.5]\),
which corresponds to the band of the leads.
The left graph in the figure (\(\epsilon_{0} = 0\)) includes the homogeneous ``bulk'' case,
namely the setup in which $t_{01}=t_L$ ($= 0.25$), i.e., all hopping parameters 
as well as the on-site energies in every layer are the same, \(\epsilon_0 = \epsilon_L\) ($=0$).
The homogeneous case, \(\epsilon_{0}=0\), \(t_{01} = 0.25\), is the point where the white lines touch the continuum.
In addition to the continuum, the spectral function displays a set of up to two high-intensity lines.
The existence of these states entails an enhancement of the spectral function, and corresponds to bound and resonance states generated by the coupling of the semi-infinite leads and the central region.  
We note that a similar distinction between bound and resonance states can be made within the single impurity Anderson model~\cite{hewson.93,vo.ro.08}.
The analysis of bound versus resonance states can also be based on the assessment of the poles of the spectral function, \cref{eq:A_00_leds}.

\begin{figure}[htbp]
  \centering
  \includegraphics{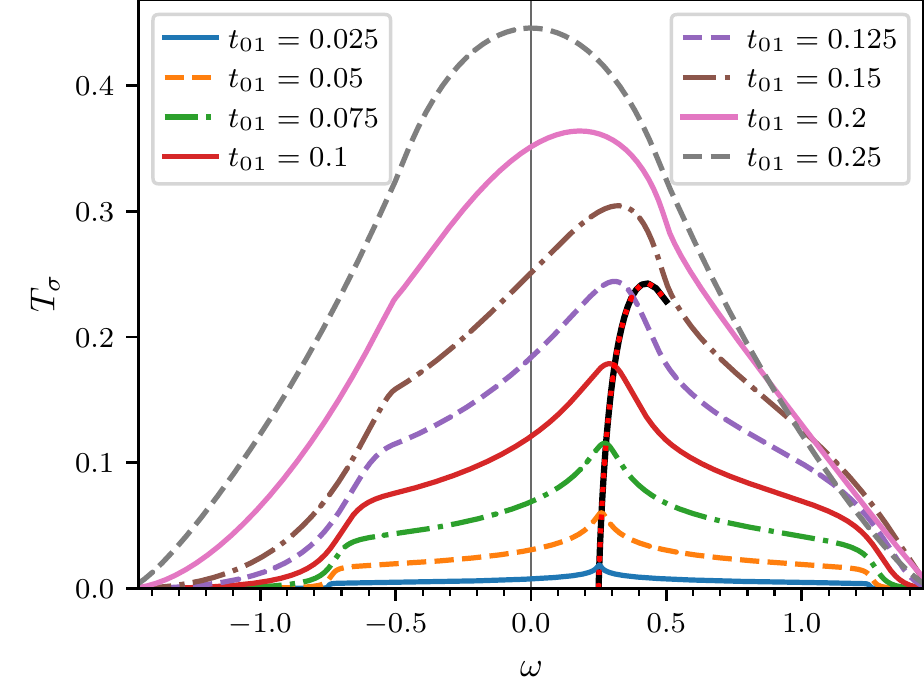}
  \caption{%
    Transmission for various values of \(t_{01}\) for \(\epsilon_{0} = 0.25\),
    corresponding to the central plot of \protect\Cref{fig:m-ferro-m_aweps_tcXX}.
    The black line with red dots is the transmission at the position of the resonance (red dotted line in \protect\Cref{fig:m-ferro-m_aweps_tcXX}).
    For small coupling, up to \(t_{01} \approx 0.1\), the maximum of transmission corresponds to the position of the resonance.
  }\label{fig:trans_U0_resonance}
\end{figure}

\begin{figure}[htbp]
  \centering
  \includegraphics{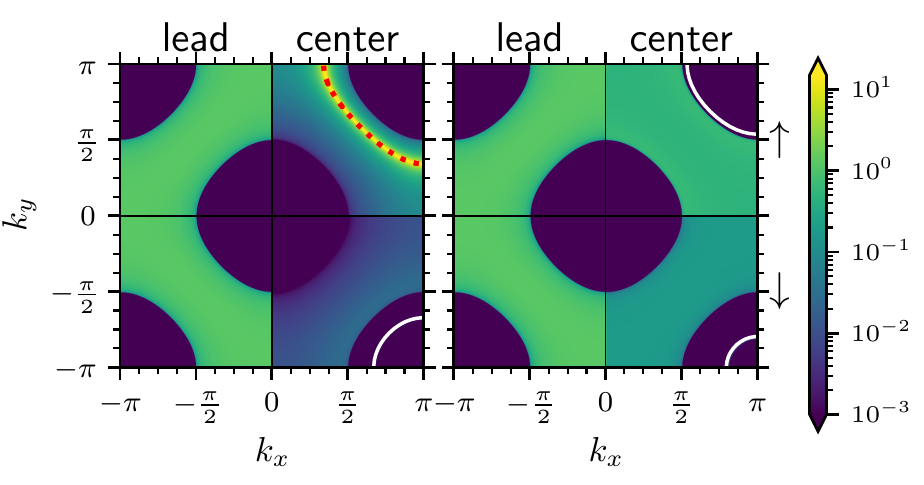}
  \caption{%
    Spectral functions of the non-interacting setup at frequency \(\omega =0\).
    The left-hand-side, \(k_{x} \in [-\pi, 0)\), of each graph shows the surface spectral function of the uncoupled lead, \(A_{L}^{s}(0-\varepsilon({k_{x}, k_{y}}))\),
    and the right-hand-side, \(k_{x} \in (0, +\pi]\), shows the spectral function of the central region, \(A_{00 \sigma}(0, \varepsilon({k_{x}, k_{y}}))\), respectively.
    The top half, \(k_{y} \in (0, + \pi]\), and 
    the bottom half, \(k_{y} \in [-\pi, 0)\), parts correspond to the respective spin directions,
    as indicated.
    Left graph: weak coupling, \(t_{01} = 0.05\);
    right graph: strong coupling, \(t_{01} = 0.25\).
  }\label{fig:m-ferro-m_U0}
\end{figure}

For larger values of the coupling, \(\abs{t_{01}}^2 > \abs{t_L}^2/2 - \epsilon_0^2/8\), bound states are located outside the continuum (white line).
Since the transmission is determined by the overlap of the spectral functions of the leads with that of the central region,
the bound states outside the continuum do not contribute to transmission.  
With decreasing $t_{01}$ values, the high-intensity states approach the continuum,
and depending on the on-site energy \(\epsilon_{0}\) they may enter the continuum region.
This leads to an enhancement of the transmission as is apparent in \cref{fig:trans_U0_resonance},
where we show the transmission for a given on-site energy of \(\epsilon_{0} = 0.25\) in the central square-lattice layer.
Up to \(t_{01} \approx 0.1\), the maximum of the transmission is given by the position of the resonance,
which is given by the black line with red dots which is \(T_{\sigma}(\omega = \Re \xi_{\sigma}(t_{01}), t_{01})\) as a function of \(t_{01}\).
Based on these findings, we can discuss the model parameters for which the investigated setup acts as an efficient
spin filter.
Clearly, choosing an on-site energy \(\epsilon_{0\sigma}\) within the continuum for one spin channel and outside the continuum for the other,
a high spin polarization of the transmission can be achieved due to the resonant states only present in one spin channel.
As an example, we consider the parameters \(\epsilon_{0}={0.5}\) and \(h_{0}={0.5}\) (\(\epsilon_{0 \uparrow} = 0.25, \epsilon_{0 \downarrow}=0.75\)).
In passing, we note that these are the same values as investigated recently~\cite{we.ot.20} for interacting bilayers.
Hence, in spite of a finite spectral weight at the Fermi level for the down-spin,
for very small values of \(t_{01}\)
we obtain a complete spin polarization in transmission, albeit with a small magnitude.
Thus, mediated by the resonant state an enhancement of the transmission spin polarization is found.
This finding is evident from \cref{fig:m-ferro-m_U0}
which shows the \(k_{\parallel}\)-resolved spectral function for the lead and the central region at \(\omega = 0\)
for a square lattice. 
The left (right) graph corresponds to \(t_{01}=0.05\) (\(t_{01}=0.25\)), while
the upper (lower) part of each graph shows the spectral function of the up-spins (down-spins),
and the left (right) part of each graph displays the lead (central region) spectral function, respectively.
The Brillouin zone (BZ) extends from \([-\pi, \pi]\); the plots are taken in one of the four (identical) quadrants of the BZ\@.
The surface spectral function of the leads, \(A^s_L(\omega^+ - \epskp)\) (left half of each graph), is identical for up- and down-spin since the leads are non-magnetic.
As the lead spectral function is calculated for the case where the lead is decoupled from the central layer, it is independent of \(t_{01}\).
The right half of each graph in \cref{fig:m-ferro-m_U0} shows the momentum resolved spectral function in the central region,
which is calculated in the presence of the leads and thus changes with \(t_{01}\).
For \(t_{01} = 0.05\),
the maximum intensity of the spectral function is located within the continuum for \(\sigma = \;\uparrow\), respectively outside the continuum for the \(\sigma = \;\downarrow\) electrons. 
In spite of a significant spectral weight in the local Green's function,
the transmission for the down-spin almost vanishes,
as the spectral weight comes from the bound states which do not contribute to the current.
Thus, we obtain a high polarization over a large frequency range.

Increasing the  coupling strength to \(t_{01} = 0.25\), we find that
the spectrum of the central region significantly changes.
The high-intensity states of the spin-up electrons are shifted out of the continuum,
becoming sharp delta peaks.
Similarly, the sharp states (white line) in the down-spin channel are repelled by the continuum and shift towards the edge of the BZ\@.
Now for both spin channels only the continuous spectrum contributes,
which is of similar magnitude for both spins.

This analysis shows that a change in the hopping amplitude between leads and central region, \(t_{01}\), significantly affects the central region spectral function and consequently modifies the transmission qualitatively, beyond a mere change of the prefactor in \cref{eq:T_omega}.
In real materials such a situation is likely to happen as electronic states are significantly influenced by structural reconstructions at the surfaces and the chemical bonding.

\subsection{Local electronic interaction in the central region}
We model the local electronic interaction in the central layer employing dynamical mean-field theory (DMFT)~\cite{me.vo.89,ge.ko.96,ko.vo.04}.
As was shown a while ago~\cite{me.vo.89}, the Hubbard model simplifies considerably in the limit of infinite spatial dimensions.
However, DMFT
provides a reliable (and non-trivial) approximation also for two and three dimensions for a large range of model parameters.
Within this approach,
the Hubbard model is self-consistently mapped onto the single-impurity Anderson model (SIAM),
 thereby allowing the use of various methods that are available for impurity problems~\cite{ge.ko.96,ko.sa.06}. 

In combination with materials-specific input, we have previously applied the DFT+DMFT~\cite{ko.sa.06} technique to heterostructures, more recently using supercells~\cite{ch.mo.15,mo.ap.17}.
While these studies were based on a perturbative impurity solver, in the present work
we employ the recently developed FTPS solver~\cite{ba.zi.17}, which is non-perturbative and allows to accurately compute spectral functions~\cite{we.ot.20}.
We emphasize that this solver works at zero temperature, and that there is no need to perform an analytic continuation of the spectral function which is otherwise a difficult technical issue~\cite{os.ch.12,os.vi.17}.
The hybridization function of the SIAM is discretized using a large number of bath sites,
in our case \(249\) sites per spin.
We calculate the ground state of the finite-size impurity problem using the density matrix renormalization group (DMRG)~\cite{whit.92,scho.05}.
Subsequently, we perform the time evolution using the time-dependent variational principle (TDVP)~\cite{ha.ci.11,lu.os.15,ha.lu.16,ba.ai.20} to obtain the local Green's function.
For the time evolution, we choose time-steps \(\Delta t = 0.1\), and a maximal time of \(t = 250\).
Presently we perform the scheme outlined above for the setup described in \Cref{sec:model}, including the Hubbard term \(U_{0}\) in the central region, see \cref{eq:H_C}.
Again, we consider the Hamiltonian with a square lattice of half-bandwidth \(D = 4t_{\parallel} = 1\) in-plane,
non-magnetic half-filled leads, \(\epsilon_{L \sigma} = 0\), with hopping \(t_{L} = 0.25\); in the central layer, we assume \(\epsilon_{0} = 0.5\) and \(h_{0} = 0.5\).
\begin{figure}[htbp]
  \centering
  \includegraphics{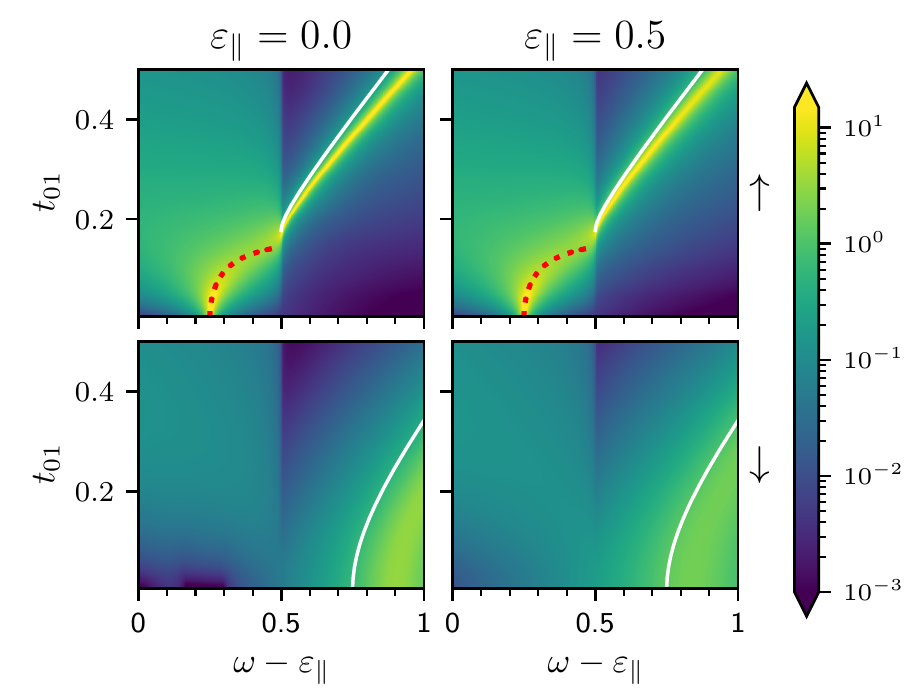}
  \caption{%
    Dependence of the spectral function of the central region, \(A_{00 \sigma}(\omega, \epsp)\), on \(t_{01}\).
    The parameters are: \(\epsilon_{L \sigma} = 0\), \(t_{L} = 0.25\), \(D=1\), \(\epsilon_{0} =0.5\), \(h_{0} =0.5\), \(U_{0} = 2\).
    The red dotted line corresponds to the real-part of equation \Cref{eq:delta-peak-solution} within the band of the leads for the \emph{non-interacting} system, \(U_{0} = 0\).
  }\label{fig:aweps_tcXX_U2}
\end{figure}
The spectral function of the central region is given by the imaginary part of \cref{eq:Gf_center} on the real axis, \(z = \omega^{+}\).
This requires the knowledge of the many-body self-energy for the various coupling strengths \(t_{01}\) for a fixed interaction strength, here \(U_{0}=2\).
We calculate the self-energy for steps of \(\Delta t_{01} = 0.05\) and linearly interpolate in-between to obtain a continuous function.
\Cref{fig:aweps_tcXX_U2} shows the spectral function.
We contrast this with the non-interacting case by plotting the delta peak outside the band (white line) and the resonance in the band (red dotted line) for \(U_{0}=0\),
which are given by the real-part of \cref{eq:delta-peak-solution}.
We observe that the resonance in the up-spin is hardly affected by the interaction.
In the regime of small \(t_{01}\) the down-spin is almost depleted,
consequently there are only small interaction effects for the up-spin.
The bound states, however, change in slope compared to the non-interacting case.
For the down-spin we observe a shift of the bound-states to higher frequencies \(\omega\),
and a considerable broadening.
\begin{figure}[htbp]
  \centering
  \includegraphics{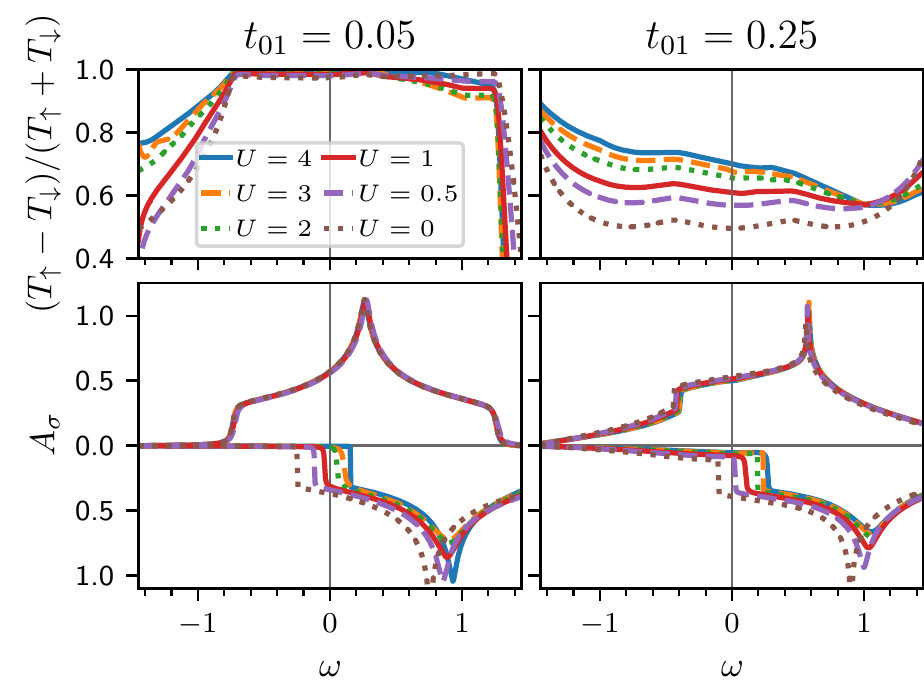}
  \caption{%
    Polarization of the transmission (top row) and local spectral function (bottom row)
    for an interacting central layer with varying on-site interaction \(U_{0}\).
    Left column: small coupling, \(t_{01} = 0.05\);
    right column: intermediate coupling, \(t_{01} = 0.25\).
    The spectral function for the down-spin varies strongly with \(U_{0}\) at the Fermi level \(\omega = 0\),
    the polarization of transmission, on the other hand, varies strongly only for \(t_{01} = 0.25\).
  }\label{fig:pol_Aw_UX}
\end{figure}
\Cref{fig:pol_Aw_UX} shows the local spectral functions of the central region, \(A_{00 \sigma}(\omega)\) (bottom), as well as the transmission polarization (top)
for weak, \(t_{01} = 0.05\) (left column), and strong, \(t_{01} = 0.25\) (right column), coupling between leads and central region, for various interaction strengths \(U_{0}\).
The local Coulomb interaction in the central region increases the polarization of the spectral function, since the minority spin states are shifted towards higher energies. 
For \(U_{0} \geq 2\)
the decoupled central region \(t_{01} = 0\) is fully polarized, which means that
it is a half-metallic ferromagnet~\cite{we.ot.20}.
While for larger values of \(t_{01}\) the leads induce states for minority electrons in the central region,
the spectral functions remain strongly polarized.
For both couplings, \(t_{01} = {0.05}\) and \(t_{01} = 0.25\),
the majority spectral function remains largely unchanged in the presence of the interaction.
This is a consequence of the almost depleted spin-down channel.
Away from the Fermi energy tails of the spectral functions are formed, which are contributing to the high-energy satellite discussed previously~\cite{we.ot.20}.
The polarization of the transmission, however, shows a behavior different from that of the local spectral function.
For \(t_{01} = 0.05\), in spite of the significant change of the minority spectral function at the Fermi energy (\(\omega = 0\)),
the polarization remains above \(95\%\) for all values of \(U_{0}\), and changes only by a few percent.
In this regime, the transmission of the majority spin is dominated by the resonance,
while the spectral weight in the minority spin derives from the bound state which does not contribute to the current.
For \(t_{01} = 0.25\), on the other hand,
the polarization of transmission follows the interaction-induced change of the local spectral function.
It increases from \(\approx 50\%\) at \(U_{0} = 0\) to \(\approx 70\%\) at \(U_{0} =4\).

\section{Conclusion}\label{sec:conclusion}
In the present work we study a one-band generic model to discuss the physics of transmission through metallic heterostructures. 
This model consists of two non-interacting leads sandwiching a central region that can be subject to local Coulomb interactions.  
The spin-dependent transmission is computed within the Meir-Wingreen formalism.
The left and right leads in our model are assumed to be identical, and the central region consists only of a single layer,
therefore the transmission decomposes into a product of the spectral function of the central region and the surface spectral function of the uncoupled leads.
Independent of the presence of the local electronic interaction,
small variations of the on-site energies and the hopping amplitude between leads and central region may strongly affect the shape of the spectral functions in the central region, and hence the conductance. 
We identify bound and resonant states that may appear in this model, depending on the coupling strength between the leads and the central region.
Generally speaking, resonance phenomena reveal themselves in the electronic conduction of mesoscopic condensed matter systems,
however, a direct characterization of the resonance energy and the line width remains non-trivial.
Bound and resonant states correspond to poles of the scattering matrix ($S$-matrix), which
relates the initial to the final state of a physical system undergoing a scattering process~\cite{ma.so.08}.
For a bound state the binding energy is directly given by a real eigenvalue, while the resonance energy and the line width are obtained from the complex eigenvalues.
By analyzing the pole structure of the Green's function corresponding to the embedded central region,
we identify the bound and the resonant states for the present setup.
As a result, for a large coupling between the leads and the central region
bound states are formed outside the continuum spectrum of the leads.
These states do not contribute to the transmission, thus we expect them to be localized.
In contrast, for the weak-coupling resonant state,
the complex poles of the Green's function entail an
enhancement of the spin polarization of the transmission.

Dynamical mean-field theory is used to solve the interacting problem in the central region.
The recently developed FTPS solver provides accurate results for the spectral functions and for the position of the bound and resonant states.
For a certain set of parameters, electronic interactions lead to an enhanced spin polarization of the spectral function,
as the minority electrons are shifted away from the Fermi level.
As a consequence, a reduction of electronic correlations for the majority spins 
is found.
For the bound states outside the continuum of the leads' spectral function, electronic correlations lead to a significant broadening.
Finally, we wish to emphasize that the results of the present study may significantly advance computational endeavors directed at the transport properties of (generally interacting) electronic systems coupled to leads.
As discussed above, for a central region sandwiched between non-interacting leads,
the effect of local electronic correlations on the resonant and bound states in the central region
can be studied by including the self-energy produced by the real-space DMFT into the Meir-Wingreen approach.
In addition, the relevance of the coupling between the leads and the central region, and the relation between bound and resonant states and the continuum beyond the non-interacting picture, certainly requires further investigations, as well as its dependence on model parameters. A particular interesting question concerns the nature of the quasiparticle states in the central region.
Furthermore, departures from the Fermi liquid description could be relevant~\cite{we.ot.20}.
The current approach also can be extended to multi-orbitals models, which are predestined to generate further challenges at the numerical level, as well as to nonlinear (e.g., finite voltage) transport properties~\cite{ap.dr.18,ec.wy.20}.

\medskip

\medskip
\textbf{Acknowledgments} \par 
Financial support by the Deutsche Forschungsgemeinschaft (project number 107745057, TRR 80) is gratefully acknowledged.

\medskip

\bibliographystyle{MSP}
\bibliography{sources}

\end{document}

%% file: commands.tex
\newcommand{\gsurf}{g^{\text{s}}} 
\newcommand{\kp}{\vb{k}_\parallel}
\newcommand{\rhop}{\rho_\parallel}
\newcommand{\epsp}{\varepsilon_{\parallel}}
\newcommand{\epskp}{\varepsilon_{\kp}}
\newcommand{\kprp}{{k_\perp}}
\newcommand{\kpl}{k_L}
\newcommand{\oc}{\hat{c}^{\vphantom{\dagger}}}
\newcommand{\ocd}{\hat{c}^\dagger}
\newcommand{\gfop}[2]{\expval*{\!\!\braket*{#1}{#2}\!}\!}

%% file: tikz/structure.tikz.tex
\makeatletter
\newcommand\sitenode[2]{site_x0_y#1_z#2}
\def\sitelist{1,2,3,4}
\def\sitelistn{2,3,4}
\def\sitelistp{1,2,3}

\tikzset{tilted/.style={%
  plane origin={(#1, 0, 0)},
  plane x={(#1, 1, 0)}, plane y={(#1, 0.25, 1)}, 
  canvas is plane,
}}
\tikzset{siteshape/.style={circle, minimum size=0.15cm, inner sep=0pt, outer sep =0pt}}
\tikzset{site/.style={shading=ball, siteshape}}
\tikzset{backgroundshadow/.style={white,very thick,opacity=0.3}}

\tikzset{coordinategrid/.pic={%
  \begin{scope}[tilted]
    \foreach \y in \sitelist{%
      \foreach \z in \sitelist{%
        \coordinate[siteshape] (\sitenode{\y}{\z}) at (\y,\z) {};
      }
    }
  \end{scope}
}}
\tikzset{pics/layer/.style args={label=#1,color=#2}{code={%
    \begin{scope}[tilted]
			\fill[fill=#2, opacity=0.6] (0, 0) rectangle (5, 5);
			\fill[fill=#2, shading=axis, left color=#2!30!white, right color=black, opacity=0.5] (0, 0) rectangle (5, 5);
      \foreach \y in \sitelist{%
        \foreach \z in \sitelist{%
          \node[site] at (#1\sitenode{\y}{\z}) {};
        }
      }
      \foreach \y in \sitelist{%
        \foreach \z [evaluate=\z as \zn using int(\z+1)] in \sitelistp{%
          \draw[semithick] (#1\sitenode{\y}{\z}) -- (#1\sitenode{\y}{\zn});
          \draw[semithick] (#1\sitenode{\z}{\y}) -- (#1\sitenode{\zn}{\y});
        }
        \draw[dashed] (#1\sitenode{\y}{1}) -- +(0, -0.5);
        \draw[dashed] (#1\sitenode{\y}{4}) -- +(0, +0.5);
        \draw[dashed] (#1\sitenode{1}{\y}) -- +(-0.5, 0);
        \draw[dashed] (#1\sitenode{4}{\y}) -- +(+0.5, 0);
      }

    \end{scope}
  }}
}
\makeatother
\newcommand{\layerandright}[3]{
  \path (#1,0) pic {layer={label=l#1,color=#3}};
  \foreach \y in \sitelist{%
    \foreach \z in \sitelist{%
      \draw[backgroundshadow] (l#1\sitenode{\y}{\z}) -- (l#2\sitenode{\y}{\z});
      \draw (l#1\sitenode{\y}{\z}) -- (l#2\sitenode{\y}{\z});
  }}
}

\begin{tikzpicture}[z=0.30cm]
  %
  %
  \foreach \lay in {-3, ..., 3}{
    \path (\lay,0) pic (l\lay) {coordinategrid};
  }
  \foreach \y in \sitelist{\foreach \z in \sitelist{%
      \draw[dashed, backgroundshadow] (l-3\sitenode{\y}{\z}) -- +(-0.5, 0);
      \draw[dashed] (l-3\sitenode{\y}{\z}) -- +(-0.5, 0);
	}}
  \foreach \lay[evaluate=\lay as \layn using int(\lay+1)] in {-3, -2, -1}{%
    \layerandright{\lay}{\layn}{green}
  }
  \layerandright{0}{1}{yellow}
  \foreach \lay[evaluate=\lay as \layn using int(\lay+1)] in {1, 2}{%
    \layerandright{\lay}{\layn}{green}
  }
  \path (3,0) pic {layer={label=l3,color=green}};
  \foreach \y in \sitelist{\foreach \z in \sitelist{%
      \draw[dashed, backgroundshadow] (l3\sitenode{\y}{\z}) -- +(0.5, 0);
      \draw[dashed] (l3\sitenode{\y}{\z}) -- +(0.5, 0);
  }}
  %
  %
  \node[below=1.7cm] at (l0\sitenode{1}{1}) {\large C};
  \begin{scope}
    \path (l1\sitenode{1}{1})++(3.0, 0) node (rightclip) {};
    \clip (l-1\sitenode{1}{1})++(-3, -2.7) rectangle (rightclip);
    \draw [decorate, decoration={brace}] (l-1\sitenode{1}{1})++(0, -1.7) -- node[below] {\large L} +(-4, 0);
    \draw [decorate, decoration={brace,mirror}] (l1\sitenode{1}{1})++(0, -1.7) -- node[below] {\large R} +(+4, 0);
  \end{scope}
  \draw[<-, thick] (l-2\sitenode{1}{1}) edge[bend left] node[below]{\contour{white}{\(t_L\)}} (l-3\sitenode{1}{1});
  \draw[<-, thick] (l-1\sitenode{1}{1}) edge[bend left] node[below]{\contour{white}{\(t_L\)}} (l-2\sitenode{1}{1});
  \draw[->, thick] (l-1\sitenode{1}{1}) edge[bend right] node[below]{\contour{white}{\(t_{-10}\)}} (l0\sitenode{1}{1});
  \draw[<-, thick] (l1\sitenode{1}{1}) edge[bend left] node[below]{\contour{white}{\(t_{01}\)}} (l0\sitenode{1}{1});
  \draw[->, thick] (l1\sitenode{1}{1}) edge[bend right] node[below]{\contour{white}{\(t_R\)}} (l2\sitenode{1}{1});
  \draw[->, thick] (l2\sitenode{1}{1}) edge[bend right] node[below]{\contour{white}{\(t_R\)}} (l3\sitenode{1}{1});

\end{tikzpicture}
  